\documentclass[a4paper,fleqn,usenatbib]{mnras}
\usepackage{newtxtext,newtxmath}
\usepackage[T1]{fontenc}
\usepackage{ae,aecompl}
\usepackage{graphicx}	
\usepackage{amsmath}	
\usepackage{amssymb}	
\usepackage[normalem]{ulem}

\title[$\ell$-changing collisions]{On the treatment of $\ell$-changing proton-hydrogen Rydberg atom collisions}

\author[D. Vrinceanu et al.]
{D. Vrinceanu,$^{1}$\thanks{E-mail: vrinceanud@tsu.edu} R. Onofrio,$^{2,3}$ and H. R. Sadeghpour$^{4}$
\\
$^{1}$ Department of Physics, Texas Southern University, Houston TX 77584, USA \\
$^{2}$ Dipartimento di Fisica e Astronomia ``Galileo Galilei", Universit\`a di Padova, Via Marzolo 8, 35131 Padova, Italy \\
$^3$ Department of Physics and Astronomy, Dartmouth College, 6127 Wilder Laboratory, Hanover NH 03755, USA\\
$^{4}$ ITAMP, Harvard-Smithsonian Center for Astrophysics, Cambridge MA 02138, USA}

\date{Accepted XXX. Received YYY; in original form ZZZ}

\pubyear{2017}

\begin{document}
\label{firstpage}
\pagerange{\pageref{firstpage}--\pageref{lastpage}}
\maketitle

\begin{abstract}
Energy-conserving, angular momentum-changing collisions between protons and highly excited Rydberg 
hydrogen atoms are important for precise understanding of atomic recombination at the photon decoupling era,
and the elemental abundance  after primordial nucleosynthesis.
Early approaches to $\ell$-changing collisions used perturbation theory for only
dipole-allowed ($\Delta \ell=\pm 1$) transitions. An exact non-perturbative quantum mechanical treatment is possible, but
it comes at computational cost for highly excited Rydberg states. In this note we show how to obtain 
a semi-classical limit that is accurate and simple, and develop further physical insights afforded by
the non-perturbative quantum mechanical treatment.
 
\end{abstract}


\section{Introduction}
The dynamics of atomic recombination and its impact on the cosmic background radiation are crucial
to constrain variants of Big Bang models \citep{Chluba2007,Chluba2010}. The recombination cascade of highly excited Rydberg
H atoms is influenced  by energy-changing \citep{Vrinceanu2014,Pohl2008} and angular momentum-changing collisional
processes (\cite{Pengelly1964} - PS64 thereafter; \cite{Vrinceanu2012} - VOS12 thereafter),
and is a major source of systematic error for an accurate determination of the recombination history.
Moreover, primordial nucleosynthesis is studied by determining the He/H abundance ratio.
This is obtained by determining the ratio of 
emission lines of He I and H I, and using the most accurate models for the recombination rate coefficients
\citep{Ferland1986,Benjamin1999,Benjamin2002,Luridiana2003,Izotov2006,Izotov2007}.

Besides cosmology, recombination rate coefficients for hydrogen and helium are also 
important in studying radio emission from nebulae \citep{Pipher1969,Brocklehurst1970,Samuelson1970,Otsuka2011}, 
and in the study of cold and ultracold laboratory plasmas \citep{Gabrielse2005}. 
In particular, there is a pending puzzle in the determination of elemental abundance and 
electron temperature in planetary nebulae, as optical recombination lines and collisionally induced 
lines provides significantly different values \citep{Izotov2006,Garcia2007,Nicholls2012,Storey2015}.

Dipole $\ell$-changing collisions $n \ell \rightarrow n \ell\pm 1$ between 
energy-degenerate states within an $n$-shell are dominant in the dynamics of proton-Rydberg hydrogen 
atom collisions, and have been addressed long ago by Pengelly and Seaton in the framework of the Bethe 
approximation in a perturbative framework (PS64). More recently, we examined (VOS12) the problem 
obtaining non-perturbative results for arbitrary $n \ell \rightarrow n \ell'$ energy-conserving transitions,
including the dipole allowed transitions, which produce rate coefficients smaller compared with PS64. 
This results in the estimation of higher densities for available spectroscopic data, which is  
of relevance at least in cosmology as different H I emissivities are derived using the two models, with 
differences of up to 10\% \citep{Guzman1}.
This in turn impacts the precision required on the primordial He/H abundance 
ratio to constrain cosmological models. 

The exact quantum expression obtained in VOS12 was complemented by a simplified classical limit
transition rate that was in good quantitative agreement with the quantum rate and also with the results of Monte Carlo
classical trajectory simulations for arbitrary $\Delta\ell$. For dipole allowed transitions, $\Delta\ell = \pm1$, 
Monte Carlo computations in VOS12 predicted a finite cross section instead of a logarithmically divergent one,
due to a discontinuity in the classical transition probability at large impact parameters.

In a recent publication \citet{Storey2015} recommended that the PS64 rates should be preferred over
the classical results in VOS12 due to how PS64
employed an {\em ad hoc} density-dependent cutoff procedure to treat the dipole-allowed angular momentum changing collisions.
In a series of papers \citet{Guzman1,Guzman2,Williams2017} investigated the influence of differently calculated $\ell$-changing rate
coefficients in CLOUDY simulations of emissivity ratios, concluding that the quantum VOS12 treatment is more appropriate when modeling
recombination through Rydberg cascades. 
In this note, we provide further validations and insights on our model and show how a slightly different classical limit is
constructed to provide non-perturbative expressions that are uniformly consistent with the quantum behavior 
for all impact parameters. In this way, the deficiency of the classical transition
rates discussed by \citet{Guzman1,Guzman2,Williams2017} is effectively eliminated.

\section{Proton-Hydrogen atom collisions at large impact parameter}
Consider an ion projectile with electric charge, in elementary units, of $Z$ moving at speed $v$ smaller or comparable with that of 
the target Rydberg electron $v_n = e^2/n\hbar$ in a state with principal quantum number 
$n$ and angular quantum number $\ell$. Results for collisions with proton are obtained by setting $Z=1$.
Even when the impact parameter $b$ is larger than the size of the Rydberg atom, $a_n = n^2 a_0$, with $n$ the 
principal quantum number and $a_0 = 0.53 \times 10^{-10}$m the Bohr radius, the weak electric field created by the 
projectile lifts the degeneracy of the Rydberg energy shell and mixes angular momentum states within the shell. 
At the end of the slow and distant collision with the ion, the Rydberg atom is in a different angular 
momentum state with the same initial energy. Therefore collisions that change angular momentum, without any 
energy transfer, have extremely large cross sections and rate coefficients. 
The rate coefficient $q$ of this process scales as $q_{n\ell \rightarrow \ell'} \sim n^4/\sqrt{T}\Delta \ell^3$ (VOS12)
with temperature $T$, and change in angular momentum $\Delta \ell = \ell' - \ell$.

Since the angular momentum changing collisions are most probable at large impact parameters it is safe to 
assume that the dipole term in the interaction energy dominates over the other multipolar contributions,
which can be therefore neglected. Moreover, as the projectile has a much greater angular momentum than
that of the target atom, it can be assumed that the projectile's angular momentum is conserved and the 
projectile moves along a straight line trajectory defined by the projectile position vector ${\bf R}(t)$. 
According to these assumptions, the Hamiltonian of the Rydberg electron contains a time-dependent interaction
potential term given by
\begin{equation}
V(t) \approx - Z e^2 \frac{{\bf r}\cdot{\bf R}(t)}{|{\bf R}(t)|^3}\quad,
\label{dipole}
\end{equation}
where ${\bf r}$ is the electron position. 
At extremely large impact parameter $b>> n^2 a_0$ the interaction potential (\ref{dipole}) may be treated 
as a perturbation and the collision can be treated in the first Born approximation for the transition probability
\begin{equation}
\begin{split}
P^{(B)}_{n\ell\rightarrow n\ell\pm1} & = \frac 1{\hbar^2} \frac 1{2\ell+1}\sum_{mm'}
\left| \int_{-\infty}^{\infty} \langle n\ell'm'|V(t)|n\ell m \rangle\;dt\right|^2 \\
& = 3 \left(\frac {Ze^2a_0}{\hbar bv}\right)^2 \frac{\ell_>}{2\ell + 1} n^2(n^2 - \ell_>^2)
\end{split}
\label{Born}
\end{equation}
where $\ell_>={\mathrm{max}}(\ell,\ell\pm1)$. This result has been obtained in the pioneering
work PS64 and forms the basis for PS64 rate coefficient 
for angular momentum changing collisions. Although simple and easy to use, the expression (\ref{Born}) 
leads to a number of severe difficulties at smaller $b$.
Various proposals were published attempting to improve the theory beyond 
the perturbation theory: close-coupling channel approximation \citep{Bray1992}, infinite level \citep{Presnyakov1970},
and rotating  frame approximations \citep{Bellomo1998}.
This also stimulated experimental investigations, in which the 
redistribution of Na(28) Rydberg atom $\ell$ states in collisions with slow Na$^+$ ions was measured \citep{Sun1993}.

Specifically, the difficulties that stem from using perturbative solutions for the potential (\ref{dipole}) are:
\begin{enumerate}
\item
The perturbative solution is derived from the matrix elements of (\ref{dipole}) with respect to
unperturbed states, and therefore only results for $\ell \rightarrow \ell \pm 1$ transitions can be obtained
with this theory, as prescribed by the dipole selection rule. 
\item 
The transition probability (\ref{Born}) diverges as $b \rightarrow 0$, violating $P_{n\ell\rightarrow n\ell\pm1} < 1$, 
reflecting unitarity. This difficulty is handled in the PS64 formulation by introducing a cutoff impact 
parameter $R_1$ such that the probability for transitions at $b \leq R_1$ is exactly $1/2$: 
$P^{(PS)}_{\ell \rightarrow \ell \pm 1}(v, b \le R_1) = 1/2$.
The justification for this adjustment was that for $b < R_1$, $P(b)$ is an oscillatory function 
with a mean value close to $1/2$. This assumption is quite reasonable for collisions involving 
energy transfer, when the cutoff $R_1$ is about the size of the atom. 
However, the probability for angular momentum changing collisions are dominated by very large
impact parameters ($b >>n^2 a_0$) and probabilities for collision at small impact parameters are much smaller than $1/2$.
In order to address this difficulty an extension to PS64 method was recently proposed \citep{Guzman2} in
which the constant $1/2$ is replaced with $1/4$ (model PS-M in that paper).
The overall trend of $P(b)$, as explained in the next sections, is to grow linearly with $b$.
This is the reason why the PS64 rates are overestimated.
\item
As $b \rightarrow \infty$, $P^{(B)}_{n\ell\rightarrow n\ell\pm1} \sim 1/b^2$, leading to a cross section
\begin{equation}
\sigma_{n\ell\rightarrow n\ell'} = 2\pi \int_0^{R_c} P_{n\ell\rightarrow n\ell'}\; bdb
\label{xSection}
\end{equation}
which diverges logarithmically as $\log(R_c)$ when the cut-off parameter $R_c\rightarrow \infty$.
The divergence of the cross section can be understood in the context of
the dynamics of degenerate quantum systems, such as the $\ell$-levels shell in a Rydberg atom.
The transition between degenerate states under the influence of a perturbation that have non-zero coupling 
matrix elements is possible no matter how weak this perturbation is.
The time scale governing transition probabilities is 
defined by the Rabi frequency, which for a degenerate system is given simply by $|V_{ab}|/\hbar$, where $V_{ab}$ 
are the transition matrix elements of the perturbation $V$ between degenerate states $a$ and $b$. 
Therefore, for weak electric fields, either produced during a very distant collision with an ion, or microfields generated by the 
surrounding plasma, the $\ell \rightarrow \ell \pm 1$ dipole transitions between Rydberg levels have rates proportional to the 
intensity of the perturbation.

\end{enumerate}

\section{Exact non-perturbative transition probability}
By taking advantage of the symmetries in the problem, an exact non-perturbative solution for the Rydberg atom dynamics under the
interaction potential (\ref{dipole}) can be obtained \citep{Vrinceanu2001a} and expressed as successive physical rotations, with 
direct interpretations both in quantum \citep{Vrinceanu2000} and classical \citep{Vrinceanu2001b} contexts.
Like in other physical situations, for example the precession of a magnetic moment in magnetic field, the source 
of similarities between quantum and classical motions is the group of symmetry operations for the given system, 
which for the hydrogen atom is $SO(4)$.

The exact result for the non-perturbative transition probability is
\begin{equation}
P_{n\ell\rightarrow n\ell'}= \frac{2 \ell' + 1}{2 j + 1}  \sum_{L = |\ell' - \ell|}^{2j}
(2 L + 1) 
\left\{ 
\arraycolsep=1pt
\begin{array}{ccc} 
\ell' & \ell & L \\ 
   j  &  j   & j 
\end{array}\right\}^2 
H_{jL}^2(\chi)
\label{prob}
\end{equation}
Here $\{\cdots\}$ is Wigner's six-$j$ symbol, and $H_{jL}$ is the generalized character function for irreducible representations 
of rotations defined by
\begin{equation}
\begin{split}
H_{jL}(\chi) & = \sum_m C^{jm}_{jmL0} e^{-2im\chi} \\
& = L! \sqrt{\frac{(2j+1)(2j-L)!}{(2j+L+1)!}} (2\sin\chi)^L C^{(L+1)}_{2j-L}(\cos\chi)
\end{split}
\label{HjL}
\end{equation}
where $C^{jm}_{j_1m_1j_2m_2}$ are the Clebsch-Gordan coefficients and $C^{(a)}_n(x)$ are ultraspherical (Gegenbauer) polynomials.
The effective rotation angle $\chi$ is
\begin{equation}
\sin\chi = \frac{2\alpha}{1+\alpha^2} 
\left[ 1 + \alpha^2 \cos \frac\pi2 \sqrt{1+\alpha^2}\right]^{1/2}
\sin \left( \frac\pi2 \sqrt{1+\alpha^2}\right)
\label{chi}
\end{equation}
with $\alpha$ a parameter that characterizes the dynamics of the ion projectile incoming at speed $v$
\begin{equation}
\alpha = \frac{3 Z n\hbar}{2 m_e v b}
\label{alpha}
\end{equation}
This parameter can be expressed as the product of the Stark precession frequency and the collision time.
Here $m_e$ is electron mass.

The probability (\ref{prob}) eliminates all the difficulties associated with the perturbative expression (\ref{Born}) as
it is not restricted to dipole transitions, it is well behaved in the $b\rightarrow 0$ limit, has simpler 
classical and semi-classical limits, as explained in the next sections, beyond the perturbative approximation.

The large $b\rightarrow \infty$ (or small $\alpha \rightarrow 0$) limit for the $\ell\rightarrow\ell\pm 1$ transition probability 
(\ref{prob}) can be obtained from the first $L=1$ term in the summation and by observing that
\begin{equation}
\lim_{\alpha\rightarrow 0} H_{j1}(\chi) = \frac{2j+1}3 \sqrt{j(j+1)} 4\alpha
\end{equation}
and that the six-$j$ symbol has a particularly simple form in this case
\begin{equation}
\left\{ 
\arraycolsep=3pt
\begin{array}{ccc} 
\ell\pm 1 & \ell & 1 \\ 
   j  &  j   & j 
\end{array}\right\}^2 = \frac{l_>(n^2-\ell_>^2)}{n (n^2-1) (4\ell_>^2 - 1)} 
\end{equation}
The result for the limit
\begin{equation}
\lim_{\alpha\rightarrow 0} P_{n\ell\rightarrow n\ell\pm 1} = \frac 43 \frac{\ell_>}{2\ell+1}(n^2-\ell_>^2)\alpha^2
\end{equation}
is identical with the perturbative result (\ref{Born}).

Equation (\ref{prob}) can be efficiently implemented for the computation of approximation-free 
transition rates for angular momentum changing collisions for use in astrophysical models,  
beyond the PS64 result. However, for $n \gtrsim 100$, the
direct summation becomes inefficient and it might lead to accumulation of truncation errors due to summation of large alternating 
sign numbers. For these cases, and also with the goal of obtaining more physics insight into this process, it is useful to investigate 
the limit $n\rightarrow\infty$ of (\ref{prob}). This can be done in two different ways, as explained in the next sections: one 
which applies for general transitions and impact parameters up to a critical value, and another one
that applies only to dipole allowed transitions and very large $b$.

\section{Classical limit}
The Bohr's correspondence principle asserts that quantum calculations tend to reproduce results obtained by using classical mechanics in
the limit of large quantum numbers. In the case of the probability (\ref{prob}) this limit is obtained by transforming the summation
into an integral and allowing quantum numbers to have continuous values, 
\begin{equation}
\label{climit}
\lim_{n\rightarrow\infty} P_{n\ell\rightarrow n\ell'} = 2\ell'n \int_0^1 
\left\{ 
\arraycolsep=1pt
\begin{array}{ccc} 
\ell' & \ell & L \\ 
   j  &  j   & j 
\end{array}\right\}^2 
H_{jL}^2(\chi)\;d(L/n)^2
\end{equation}

The classical limit of Wigner's six-$j$ symbol \citep{Ponzano1968} is given by $1/24\pi\sqrt{V_T}$ in terms of the volume $V_T$ of 
a tetrahedron made by the six angular momentum quantum numbers. By using the Cayley-Menger determinant to calculate this volume, 
one gets for arbitrary
transitions that
\begin{equation}
\label{c6j}
\begin{split}
& \lim_{n\rightarrow\infty \atop L/n < \infty}
\pi n^3 \left\{ 
\arraycolsep=1pt
\begin{array}{ccc} 
\ell' & \ell & L \\ 
   j  &  j   & j 
\end{array}\right\}^2  =
\lim_{n\rightarrow\infty \atop L/n < \infty} \left (2/n^6
\left|
\arraycolsep=1pt
\begin{array}{ccccc}
0 &   1 & 1   & 1   & 1\\
1 &  0  & j^2 & j^2 & j^2\\
1 & j^2 &  0  & \ell^2 & {\ell'}^2\\
1 & j^2 & \ell^2 & 0 & L^2\\
1 & j^2 & {\ell'}^2 & L^2 & 0\\
\end{array}
\right|\right)^{-1/2} \\
&  = \frac 1{\sqrt{\sin(\eta_1 + \eta_2)^2 - (L/n)^2}} \frac 1{\sqrt{(L/n)^2 - \sin(\eta_1 - \eta_2)^2}}
\end{split}
\end{equation}
Here the limit is taken such that the ratio $L/n$ remains finite, as well as the ratios for the initial and final 
angular momenta defined through $\cos\eta_1 = \ell/n$ and $\cos\eta_2 = \ell'/n$. This classical limit is
valid only for values that make the arguments of the square root positive, which limits
the integration range in $L/n$. For example, $L/n > \sin(\eta_1 - \eta_2)$, which depends on the change
$\Delta\ell$ of angular momentum in transition.

The generalized character function $H_{jL}$ is the solution of a differential equation that can be interpreted
as Schr\"odinger's equation for a particle confined by a $1/\sin^2 \chi$ potential well, that has infinite barriers
at $\chi=0$ and $\chi=\pi$ and a minimum at $\chi=\pi/2$. A WKB approximation for this problem is obtained as
\begin{equation}
\label{cHjL}
\lim_{n\rightarrow\infty\atop L/n < \infty}H_{jL}(\chi) = \frac 1{\sqrt{2\sin\chi}} (\sin^2\chi - (L/n)^2)^{-1/4}
\end{equation}
and is in excellent agreement with the exact solution at any $\chi$, except at the classical turning points ($|\sin\chi| = L/n$) where
the WKB approximation diverges, showing that classically the particle tends to be found with infinite probability at the turning  points.
Beyond the turning points, the classical probability is zero while the exact solution decreases to zero gradually. This contradictory
behavior is characteristic to the WKB approximation, and leads in the present case to a discontinuity in the transition probability
as a function of $b$, as shown in Figures 1 and 2. The nature of this discontinuity is discussed below.

Figure 1 demonstrates graphically that probability (\ref{prob}) converges to (\ref{climit}) in the $n\rightarrow \infty$ limit,
showing a linear increase up to a maximum impact parameter, followed by a sharp drop.
\begin{figure}
\includegraphics[width=3.4in]{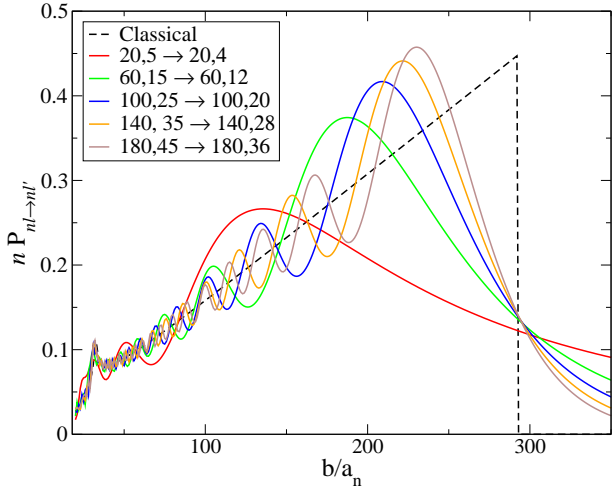}
\caption{The convergence of quantum results toward the semiclassical limit, as expected 
from the correspondence principle. The parameters are chosen such that the ratios $\ell/n$ and
$\ell'/n$ are preserved in all examples. The probabilities are also scaled by $n$ to 
obtain the semi-classical limit which obeys the classical scaling. Here $a_n = n^2 a_0$.}
\label{convergence}
\end{figure}

By combining equations (\ref{c6j}) and (\ref{cHjL}), we see that classical probability is nonzero only when 
$\sin\chi < L/n < |\sin(\eta_1 -\eta_2|$. Otherwise, integration (\ref{climit}) has analytic results in terms 
of elliptical integrals (see VOS12 for details). It is interesting to note that the same result 
was obtained directly from the classical solution of the motion under potential (\ref{dipole}) and by defining
the transition probabilities as ratios of phase space volumes \citep{Vrinceanu2000}. 
The resulting classical limit agrees very well with the non-perturbative result (\ref{prob}) as seen 
in the inset in Fig 2, for 
all $b$, except at very large $b$, where the probability drops to zero abruptly, instead of showing the $1/b^2$ decay of (\ref{Born}).

For $1<<b< b_{max}$, which means small $\alpha$ and $\chi$, only small angular momentum changes are possible and 
one can approximate $\sin\chi \approx 2\alpha$, $\sin(\eta_1 - \eta_2) \approx \Delta \ell/\sqrt{n^2-\ell^2}$ and
$\sin(\eta_1+\eta_2) \approx 2\ell\sqrt{n^2-\ell^2}/n^2$, to provide a much simplified transition probability
\begin{equation}
\label{cprob}
P^{(C)}_{n\ell\rightarrow n\ell'} = \left\{ 
\begin{array}{cc}
b/2b_{max} & \mbox{ for } b \le b_{max}/\Delta\ell \\
0          & \mbox{ for } b >   b_{max}/\Delta\ell
\end{array}
\right.
\end{equation}
where the classical cutoff radius $b_{max} = 3n a_0 \sqrt{n^2 - \ell^2} Ze^2/\hbar v$ is obtained from
the cusp relation $\sin\chi = |\sin(\eta_1 - \eta_2)|$. This linear increase for $b<b_{max}$ is in contrast with the {\it ad-hoc} PS64 assumption that the probability is 1/2 for $b<R_1$, and it explains why the PS64  rate coefficient is larger than the quantum VOS12 rate coefficient.

The abrupt discontinuity in $b$ at $b_{max}$ displayed by equation (\ref{cprob}) is problematic, reflecting the deficiency 
of the WKB approximation to describe quantum tunneling. The most significant difficulty for (\ref{cprob}) is for dipole 
allowed $|\Delta\ell|=1$ transitions that have logarithmically divergent cross sections. Instead, by using
probability (\ref{cprob}) in integrating (\ref{xSection}) the result is a finite cross section, denoted as $\sigma_C$ for
future reference.
For all other $|\Delta\ell| > 1$ transitions, the sharp discontinuity has a minor effect since both the classical and 
quantum transitions have finite cross sections and rate coefficients, and the approximation (\ref{cprob}) works surprisingly well.
The next section shows how to address the deficiency of classical probability (\ref{cprob}) for $\Delta\ell=1$ at 
$b = b_{max}$ by taking the classical limit differently. This procedure is akin to the textbook prescription of 
treating the WKB singularity at the turning points, by developing a local approximation around those points and 
then "stitching" together approximations over various intervals.

\section{Semiclassical limit}
Instead of the classical approximation (\ref{cHjL}) valid over a wide range of $\chi$ values, we use a local approximation
\citep{Varshalovich}
\begin{equation}
\lim_{n\rightarrow\infty\atop \alpha\rightarrow 0, \alpha n < \infty} \frac 1n H_{jL}(\chi) = j_L(2 \alpha n)
\end{equation}
valid only for small $\alpha$, as long as the product $\alpha n$ is finite.
Here $j_L(x)$ is the spherical Bessel function.
 
By using this approximation in the integration (\ref{climit}), and working only for dipole transitions $\ell'=\ell\pm1$, we
obtain a semiclassical transition probability as the integral
\begin{equation}
P^{(SC)} = \frac{2\ell}{\pi} \int_1^n \frac{j_L^2(n\alpha) \;dL}{\sqrt{4\ell^2[1 - (\ell/n)^2] - L^2}}
\end{equation}
which is dominated by values around the $L=1$ end of the integration range. Since $j_1(x) \approx x/3 + {\cal O}(x^3)$,
this semiclassical transition probability has the correct asymptotic $\sim 1/b^2$ at $b\rightarrow \infty$ limit. The integral
can be approximated to get
\begin{equation}
\label{scprob}
P^{(SC)} \approx \frac 32 j_1^2( 2 \alpha\sqrt{n^2-\ell^2})
\end{equation}

\begin{figure}
\centerline{\includegraphics[width=3.3in]{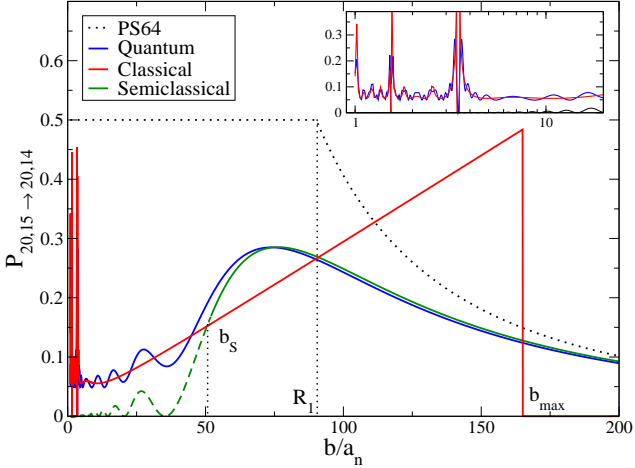}}
\caption{Probability for transitions within the $n=20$ hydrogenic shell from $\ell=15$ to $\ell'=14$ in collisions with 
protons having speed $v=0.25 v_n$ as a function of scaled impact parameter. The quantum theory is contrasted with the classical 
and semiclassical approximations and with the perturbative result in Equation (\ref{Born}).
We observe that the PS-M model \citep{Guzman2} brings the results to better agreement with quantum results than PS64.
The inset shows the good agreement between the classical approximation and the quantum result at small impact parameter.
}
\label{probs}
\end{figure}
Figure 2 shows the PS64 perturbation theory (\ref{Born}), classical approximation (\ref{cprob}) and semiclassical approximation
(\ref{scprob}) for a dipole allowed transition as compared with the quantum probability (\ref{prob}). The classical limit agrees
well with the exact result for low and moderate impact parameters (as shown in inset),
displaying the abrupt classical discontinuity at $b_{max}$.
On the other hand, the semiclassical approximation does well at very large $b$, 
but fails at small $b < b_S$, as shown in the figure by a dashed line.

In order to take advantage of the good agreement of the classical and semiclassical transition probabilities in their respective 
ranges and obtain an accurate approximation for the cross section, we combine them in an {\em effective} transition probability defined as:
\begin{equation}
\label{eprob}
P^{(E)}_{n\ell\rightarrow n\ell'} = \left\{ 
\begin{array}{cc}
b/2b_{max}  & \mbox{ for } b \le b_{S} \\
\frac 32 j_1^2( b_{max}/b )    & \mbox{ for } b >   b_{S}
\end{array}
\right.
\end{equation}
with the matching $b_S = \gamma b_{max}$ defined as the smallest impact parameter for which the classical and semiclassical 
approximations are equal, ensuring the continuity of the probability, and $\gamma = 0.3235133$ is the solution to the 
transcendental equation $j_1^2(1/x) = x/3$. 

\begin{figure}
\includegraphics[width=3.4in]{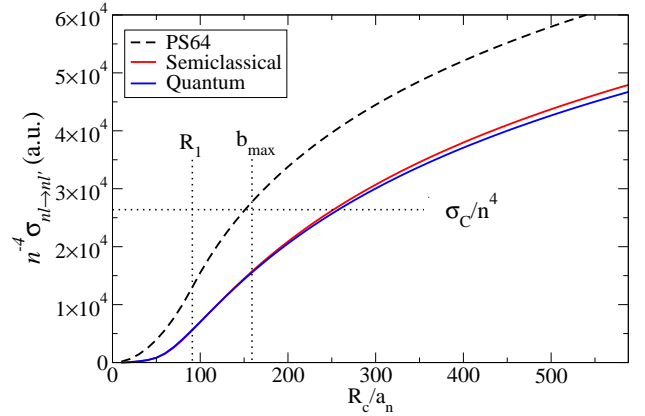}
\caption{The cumulative cross section $\sigma_{\ell\rightarrow\ell'}$ in atomic units as a function of the scaled cutoff 
parameter $R_c$ for the exact quantum theory, its semiclassical limit and for the PS64 perturbative approximation. 
The plot shows the (20,15)$\rightarrow$(20,14) transition in collisions with ions with speed $v = 0.25 v_n$.
The corresponding scaled finite classical cross section $\sigma_C$ is marked on the graph.
The low $b$ cutoff $R_1$ used by PS64 and the $b_{max}$ impact parameter after which the classical transition
probability is zero, are shown as dotted lines.}
\label{cumulative}
\end{figure}
The cross section is calculated by using Eq. (\ref{xSection}) to get the semiclassical cross section
\begin{equation}
\label{sccs}
\sigma^{(SC)}_{n\ell\rightarrow n\ell'} = \frac{\pi b_{max}^2}{3}
\left\{\begin{array}{ll}
(R_c/b_{max}^2)^3 & , R_c \le b_S \\
\gamma^3 + \left[ T(R_c/b_{max}) - T(\gamma)\right] & , R_c > b_S
\end{array}\right.
\end{equation}
where the function $T$ is 
\begin{equation}
\begin{split}
T(x) = & -Ci(2/x) + 3*x^4(3+2x^2)/8 - \\
& x^2(2-3x^2 + 6x^4)\cos(2/x)/8 + \\
& x(2-x^2-6x^4) \sin(2/x)/4
\end{split}
\end{equation}
and $Ci(z) = - \int_z^\infty \cos(t)/t\; dt$ is the cosine integral function.

Figure 3 shows calculations of the cumulative 
transition cross section as a function of the cutoff parameter $R_c$ used to regularize the logarithmic singularity.
The PS64 result overestimates the non-perturbative quantum cross section
derived from Eq. (\ref{prob}) by amounts that depend on the cutoff parameter $R_c$.
As explained in section 2, the PS64 rates are overestimated because the
probability of transition is assumed to be 1/2 for $0 < b < R_1$, while the non-perturbative calculation demonstrates
that the probability increases linearly with $b$.
Asymptotically, both PS64 and the semiclassical cross sections (\ref{sccs}) diverge logarithmically as 
$\sim \mbox{const} + \pi b_{max}^2 \ln(R_c)/3$ with $R_c \rightarrow \infty$, but with the PS64 constant
approximately twice as large as the semiclassical one.
Therefore, even for high temperature and density considered by \citet{Guzman1,Guzman2} the PS64 rate
overestimates the $\ell$-changing rate by a constant amount. This difference is independent of $R_c$, and therefore
the ratio of the two rates approaches unity in the $R_c\rightarrow\infty$.
The PS-M model also has the linear increase with $b$ and the same asymptotic behavior,
but as noted in their paper, the agreement with the quantum VOS12 model is reasonable good in general,
similar with the results derived from Eq. (\ref{eprob}), but deficient in some
extreme cases, such as low $\ell$ values.

Recent papers \citep{Guzman1,Guzman2,Williams2017} argued that quantum formula (\ref{prob}) is
computationally expensive, while the classical limit (\ref{cprob}) has an abrupt drop, instead of the $1/b^2$ decay
as $b\rightarrow\infty$, and therefore the PS64 perturbative rates should be still preferable. Figure 3 addresses this
concern by showing that semiclassical cross sections, and by extension the transition rate coefficients, are
consistent with quantum non-perturbative results, but easier to use in practical calculations due to the simplicity of the effective probability (\ref{eprob}). 

\section{Conclusions}
We have contrasted two different models for the evaluation of proton-Rydberg atom angular changing 
collision, with particular emphasis on the anatomy of their assumptions and approximations, and the 
comparison to the full quantum-mechanical setting at small principal quantum numbers.
We argue that parameters of astrophysical interest derived from diverging cross-sections 
contain a degree of arbitrariness in principle reflected in large and unknown systematic errors. 
In the absence of full quantum calculations or of precision laboratory measurements, it is more meaningful to use models with clearer physical interpretation, less assumptions, and controllable approximations. 
We believe that this pluralistic approach is even more imperative in astrophysics, since the models 
involved in the extraction of astrophysical parameters from observations are typically the major source 
of systematic error, as already extensively advocated in \citet{Mash1,Mash2,Berg,Hill}.

It was advocated in \citet{Guzman1,Guzman2} that VOS12 quantum rates to be used when high accuracies are required and faster PS64 when that accuracy is not needed to speed the calculations.
The results introduced here, derived from improved semiclassical limit (\ref{eprob}), are accurate over the whole range of impact parameters and computationally inexpensive, eliminating the dilemma of having to choose speed over accuracy.

\section*{Acknowledgments}
This work was supported by the National Science Foundation through a grant to ITAMP at the Harvard-Smithsonian Center for Astrophysics.
One of the authors (DV) is also grateful for the support received from the National Science Foundation through grants for the Center for Research on Complex Networks (HRD- 1137732), and Research Infrastructure for Science and Engineering (RISE) (HRD-1345173).
We thank G. Ferland, and his collaborators, for fruitful and stimulating dialog on this topic.

\bsp	
\label{lastpage}
\end{document}